\documentclass[aps,preprint,showkeys,preprintnumbers,amsmath,amssymb,floatfix,nofootinbib,prd]{revtex4-1}
\usepackage{hyperref}
\usepackage{color}   
\usepackage{graphicx}
\usepackage{dcolumn}

\begin{document}

\def\bb    #1{\hbox{\boldmath${#1}$}}

\title{Inelasticity resulting from rapidity spectra analysis}

\author{Zbigniew W\l odarczyk}
\email{zbigniew.wlodarczyk@ujk.edu.pl}
\affiliation{Institute of Physics, Jan Kochanowski University, 25-406 Kielce, Poland}

\author{Maciej Rybczy\'nski}
\email{maciej.rybczynski@ujk.edu.pl}
\affiliation{Institute of Physics, Jan Kochanowski University, 25-406 Kielce, Poland}

\begin{abstract}
In this work we study the pseudorapidity spectra o charged particles produced in proton+proton and proton+antiproton interactions in a wide energy range using the non-extensive Tsallis approach. We evaluate the inelasticity coefficients of the discussed reactions which remain approximately independent of the collision energy.
\end{abstract}
\date{\today}

\maketitle

\section{Introduction}
\label{sec:intro}

The concept of inelasticity in proton+proton interactions:
\begin{equation}
K=\int_{0}^{1}\frac{x}{\sigma}\frac{d\sigma}{dx}=\frac{W}{\sqrt{s}},
\label{eq:inelasticity}
\end{equation}
where $x$ is the Feynman variable, $\sigma$ is the cross-section and $\sqrt{s}$ is the center-of-mass energy of colliding protons, defines the amount of energy $W$ used effectively for the production of secondaries and plays an important role in understanding of multiparticle production processes. It says us that in the collisions only a fraction $W=K\sqrt{s}$ of the whole invariant energy $\sqrt{s}$ is spent for production of new particles while the rest of it is taken by leading particles to the forward and backward phase-space regions. Inelasticity $K$ as a dynamical variable was introduced for the first time in cosmic rays experiments,~\footnote{\textit{Leading particle effect} was discovered in the late 1940s by Zatsepin in his \textit{nuclear cascade process}~\cite{zatsepin,ginzburg}. Among others, Cocconi also discussed extensively the inelasticity already in his 1958 model for nucleon-nucleon collisions~\cite{Cocconi:1958zz}.} where it is crucial for the understanding and proper interpretation of the development of cosmic ray cascades~\cite{zatsepin,ginzburg,Cocconi:1958zz}. With an increase of energies attainable in accelerators, experimentalists rediscovered the importance of the leading particle effect~\cite{Basile:1980ap,Basile:1980by,Basile:1982we}. The importance of the ISR results obtained by the Bologna-CERN-Frascati Collaboration with respect to the \textit{effective energy} and its pendant, the \textit{leading particle effect} was rightly emphasized by Zichichi~\cite{zichichi}.

Apart from cosmic ray physics, the notion of inelasticity is natural in statistical and hydrodynamical models of multiparticle production in which $K$ is the main and essential input~\cite{Carruthers:1983xk,Shuryak:1984nq}. It also seems justified to expect these concepts to be explained by non-perturbative QCD. As a matter of fact, the inelasticity and in particular its energy dependence has been by now the subject of several QCD-inspired theoretical studies~\cite{Fowler:1984ps,Fowler:1985dv,Fowler:1987pt,Fowler:1990mf}. The energy dependence of inelasticity has a long story. While initially the inelasticity was introduced as a constant parameter, $K\cong 0.5$, later it was found in cosmic ray and accelerator experiments that $K$ decrease with energy~\cite{Wlodarczyk:1995eu}~\footnote{Recently decreasing inelasticity was advocated in Ref.~\cite{Beggio:2020dry}}. In models the energy dependence of inelasticity is an open problem. The decrease of inelasticity with energy was advocated by some authors while the others proposed that inelasticity in an increasing function of energy~\cite{Wlodarczyk:1995eu,Shabelski:1991mm}.

In this paper, the inelasticity in proton+proton (proton+antiproton) collisions in the energy range 17 GeV - 8000 GeV has been estimated from rapidity spectra analysis. The article is organized as follows. In Sec.~\ref{sec:met} we briefly describe the methodology used for description of rapidity spectra, and discuss the resultant fits to them. Sec.~\ref{sec:res} describes the procedure of extraction of inelasticity coefficient from the rapidity spectra together with discussion of the energy dependence of the obtained inelasticities. Finally, in Sec. \ref{sec:cr} we come to our summary.

\section{Pseudorapidity distributions}
\label{sec:met}

In description of multiparticle production processes one often uses statistical methods and concepts which follow the classical Boltzmann-Gibbs (BG) approach. However, as demonstrated recently, to account for some intrinsic fluctuations in the hadronizing system one should rather use the non-extensive Tsallis statistics~\cite{Tsallis:1987eu,Tsallis:2008mc,Tsallisbook}, in which one new parameter $q$ describes summarily the possible departure from the usual BG case (which is recoverd in the $q\rightarrow 1$ limit)~\cite{Wilk:1999dr}. Here we shall provide detailed description of the pseudorapidity spectra of particles (mostly pions) produced in proton+proton (proton+antiproton) collisions using Tsallis statistics approach to this problem and in this way accounting for their non-equilibrium character.

Distribution of the energy $E$ of secondary particles produced in collisions may be described by the well-known Tsallis non-extensive formula~\cite{Tsallis:1987eu,Tsallis:2008mc,Tsallisbook}~\footnote{For an  updated  bibliography  on  this  subject,  see {\tt http://tsallis.cat.cbpf.br/biblio.htm}}:
\begin{equation}
E\frac{dN}{dE}=\exp_{q}{\left(-\frac{E}{T}\right)}=\left[1+\left(q-1\right)\frac{E}{T}\right]^{1/\left(1-q\right)}
\label{eq:Tsall_dist}
\end{equation}
with parameter $T$ denoting the {\it longitudinal temperature} of the system. The $\exp_{q}{\left(x\right)}$ in the limit of $q\rightarrow 1$ becomes standard exponential function:
$\lim_{q \to 1} \exp_{q}\left(x\right)=\exp\left(x\right)$.

The energy of produced particle may be expressed in terms of particle rapidity, $y$:
\begin{equation}
E\left(y\right)=\mu_{T}\cosh{y},
\label{eq:E_rap}
\end{equation}
where $\mu_{T}=\sqrt{p_{T}^{2}+m^{2}}$ is particle transverse mass expressed by its transverse momentum $p_{T}$ and mass $m$~\footnote{For the purpose of the present study we always assume pion mass, $m=0.14$~GeV/c$^{2}$ for all produced particles.}. Thus,
\begin{equation}
dN=\exp_{q}{\left(-\frac{E\left(y\right)}{T}\right)}\beta dy,
\label{eq:dN_y}
\end{equation}
where $\beta=p/E$ denotes velocity of the given particle ($p$ is the particle momentum). Due to some problems with particle identification many high-energy physics experiments focuses on measurements of particle's \textit{pseudorapidity} instead of rapidity. Thus it is sometimes convenient to express particle production in terms of pseudorapidity, $\eta$~\cite{Liu:2013xba}. Then, the particle energy equals:
\begin{equation}
E\left(\eta\right)=\sqrt{m^{2}+p_{T}^{2}\cosh^{2}{\eta}},
\label{eq:E_eta}
\end{equation}
and the particle velocity is:
\begin{equation}
\beta\left(\eta\right)=\frac{\cosh{\eta}}{\sqrt{\cosh^{2}{\eta}+m^{2}p_{T}^{-2}}}.
\label{eq:beta_eta}
\end{equation}
Using $dy=\beta d\eta$ in Eq.~\ref{eq:dN_y} we have
\begin{equation}
dN=\exp_{q}{\left(-\frac{E\left(\eta\right)}{T}\right)}\beta^{2}\left(\eta\right) d\eta.
\label{eq:dN_eta}
\end{equation}
In the approach presented in this paper we substitute the transverse momentum of secondary particles by their average transverse momentum, $\langle p_{T}\rangle$ which depends on the energy of colliding protons~\cite{Khachatryan:2010us}:
\begin{equation}
\langle p_T\rangle\left(s\right)=A + B\ln s + C\ln^{2} s,
\label{eq:pt_s}
\end{equation}
where $A=0.413$, $B=-0.0171$, and $C=0.00143$.

Multiplicity distributions of secondary particles may be well described by the negative binomial distribution. The non-extensivity parameter $q$ is strongly connected with fluctuations of multiplicities~\cite{Wilk:2006vp}:
\begin{equation}
q-1=\frac{1}{k}=\frac{{\rm Var}\left(N\right)}{\langle N\rangle^{2}}-\frac{1}{\langle N\rangle}.
\label{eq:k_q}
\end{equation}
The negative binomial distribution shape parameter $k$ is always positive (thus $q>1$) and depends on the energy of colliding protons as~\cite{GeichGimbel:1987xy}:
\begin{equation}
k^{-1}=a + b\ln \sqrt{s},
\label{eq:k_param}
\end{equation}
where $a=-0.104\pm 0.004$ and $b=0.058\pm 0.001$.
\begin{figure}
\begin{center}
\includegraphics[width=0.98 \textwidth]{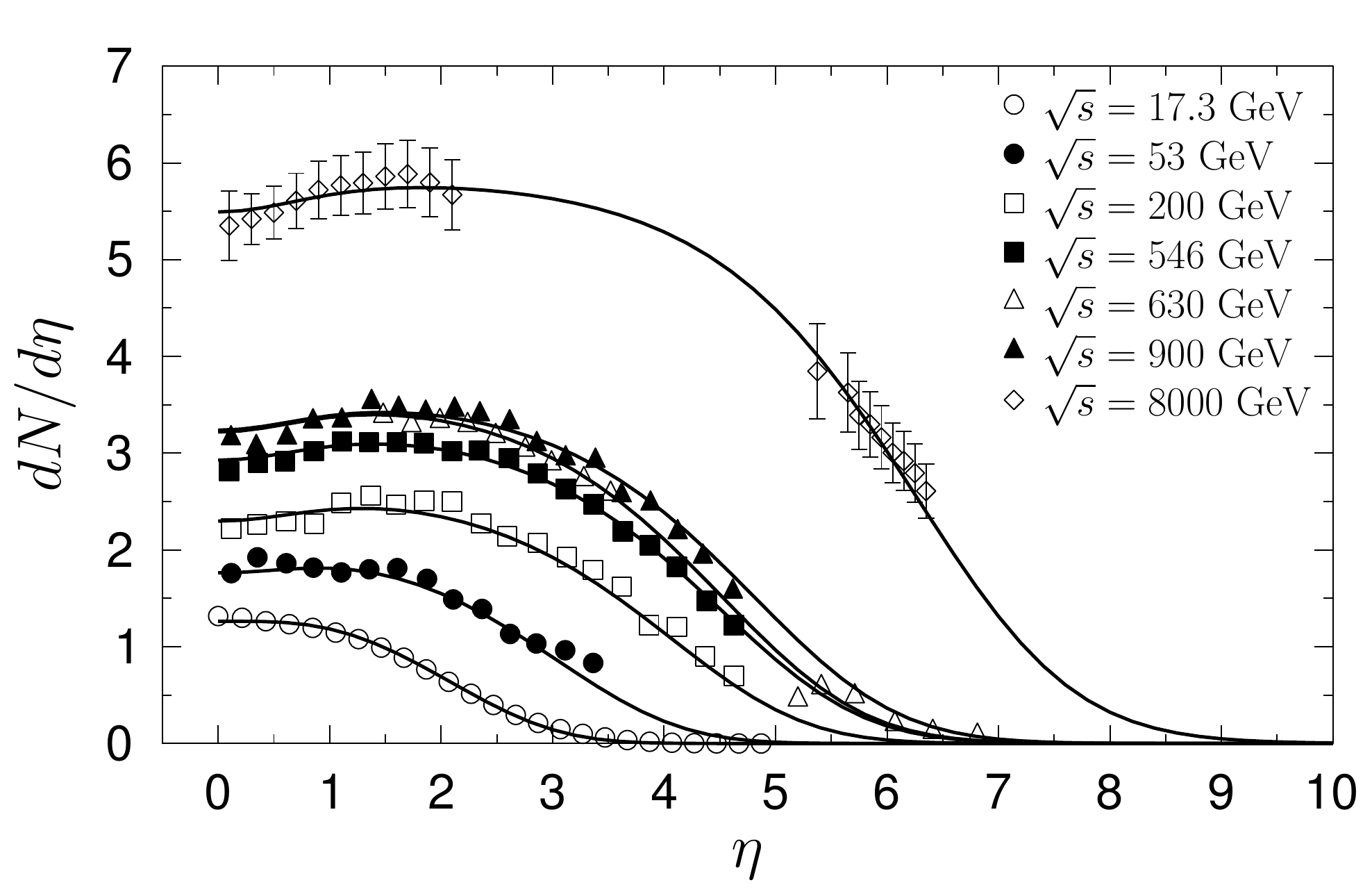}
\end{center}
\vspace{-5mm}
\caption{Pseudorapidity distributions of charged hadrons produced in proton+proton (proton+antiproton) interactions and registered by the NA49~\cite{Alt:2005zq}, UA5~\cite{Alner:1986xu}, UA7~\cite{Pare:1989mr}, and CMS and TOTEM~\cite{Chatrchyan:2014qka} experiments. With lines we show our fits obtained using Eq.~\ref{eq:dN_eta}.} 
\label{fig:eta_fits}
\end{figure}

We show in Fig.~\ref{fig:eta_fits} our results of fitting the experimental data on pseudorapidity distributions of secondaries produced in proton+proton (proton+antiproton) collisions~\cite{Alt:2005zq,Alner:1986xu,Pare:1989mr,Chatrchyan:2014qka} by using formula~\ref{eq:dN_eta}. As seen in Fig.~\ref{fig:eta_fits}, a very good agreement with data has been obtained with apparently three parameters: the ``longitudinal temperature'' $T$, mean transverse mass $\mu_{T}$ and parameter $q$ - all dependent on the energy of reaction $\sqrt{s}$. However, after closer inspection it turns out that parameter $q$, which can be regarded as a measure of fluctuations existing in the physical system under consideration, follows essentially the fluctuations of multiplicity of particles produced at given energy. Similarly, parameter $\mu_{T}$ is closely connected with the average transverse momentum, $\langle p_{T}\rangle$ known from experimental data. This shows that the only parameter which is entirely ``free'' is the {\it longitudinal temperature} $T$. Therefore we can say that what we are proposing here is essentially one-parameter fit successfully describing data on pseudorapidity distributions.

The concept of longitudinal temperature has a long history~\cite{rafelski_book}. If we decide to treat the longitudinal and the transverse motion independently, there is no reason to insist that the two temperatures associated with these motions should be equal. Such a picture would rather naturally leads to an effective longitudinal temperature $T$ being about $\gamma=\sqrt{s}/\left(2\cdot M_{p}\right)$ ($M_{p}$ denotes proton mass) times larger than transverse temperature $T_{\bot}$. We only have to assume that, in most cases, the collision time is so short that a thermal equilibrium ($T\cong T_{\bot}$) cannot be reached. The simple model described in~\cite{Hagedorn:1965fie} shows that the fraction of collisions in which thermal equilibrium may be reached, is about $1/\gamma^{2}$ of all inelastic ones, i.e., in nearly all collisions the thermal equilibrium is {\it not} reached and the longitudinal and transverse temperatures can be different. Then the above treatment would be justified.
\begin{figure}
\begin{center}
\includegraphics[width=0.98 \textwidth]{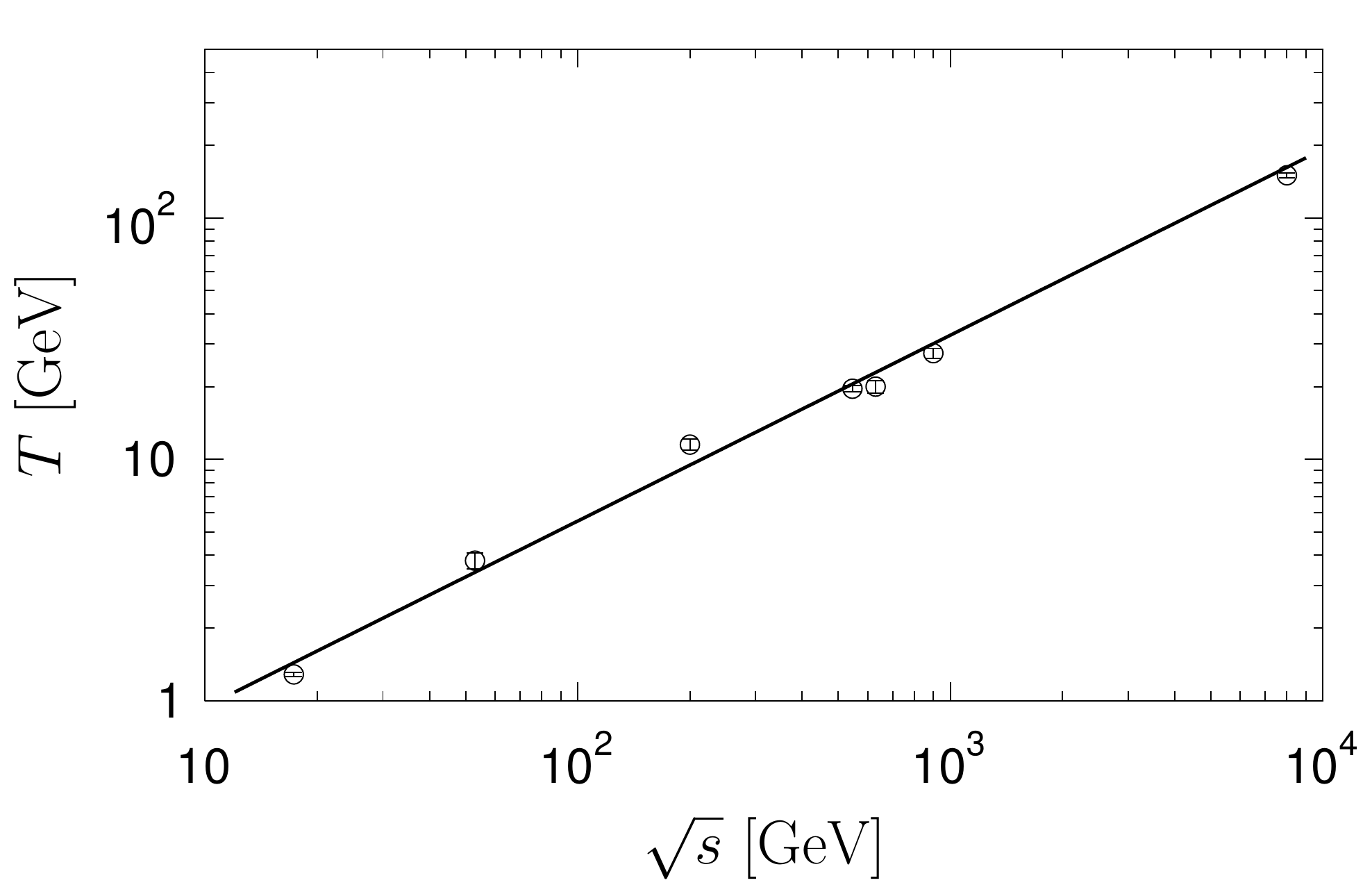} 
\end{center}
\vspace{-5mm}
\caption{Energy dependence of the longitudinal temperature parameter extracted from the fits presented in Fig.~\ref{fig:eta_fits}. With the line we show fit to the points done using Eq.~\ref{eq:T_fit}.} 
\label{fig:T}
\end{figure} 
In Fig.~\ref{fig:T} we show the evaluated energy dependence of longitudinal temperature. In discussing the energy dependence of longitudinal temperature we (following the emitting center model~\cite{Ohsawa:2019mii}) consider emitting source aligned on the rapidity axis.

In the nucleons collision a fireball is created at the collision point through the deposited energy by nucleons. The fireball is made of a hadronic gas of high initial temperature, $T_{0}=K\sqrt{s}/2$ and expands along the collision axis. The temperature of the gas decreases through the expansion and the constituent particles transmute into the produced hadrons of mainly pions ({\it hadronization}) when the density of the constituent particles of the gas arrives at a certain fixed value. Assuming adiabatic expansion of the gas, the temperature at hadronization is
\begin{equation}
T=T_{0}\left(\frac{4\pi R_{0}^{3}/3}{\pi R_{0}^{2}L}\right)^{\kappa-1},
\end{equation}
where $R_{0}$ is the nucleon radius and $\kappa=c_{p}/c_{V}$ is the specific heat ratio. The length of the gas (of a cylindrical shape) in the final state, $L=R_{0}\xi$ is assumed to be proportional to the average multiplicity $\langle N\left(\sqrt{s}\right)\rangle$, i.e. $\xi=\frac{4}{3}\xi_{0}\langle N\rangle$. The evaluated energy dependence of multiplicity:
\begin{equation}
N\left(\sqrt{s}\right)=2.73\left(\sqrt{s}\right)^{0.36}
\end{equation}
and temperature
\begin{equation}
T\left(\sqrt{s}\right)=0.16\left(\sqrt{s}\right)^{0.77}
\label{eq:T_fit}
\end{equation}
lead to $\frac{4}{3}\xi_{0}=0.98$ and $\kappa=1.64$. The value of the specific heat ratio is near to $\kappa=5/3$ for a mono-atomic gas in a classical limit~\cite{Faruk}.

\section{Inelasticity}
\label{sec:res}

The inelasticity $K$ of hadronic reactions, understood as the fraction of the incident beam energy used for the production of secondary particles can be calculated as~\cite{Navarra:2003am}:
\begin{equation}
K=\frac{2}{\sqrt{s}}\int_{0}^{\eta_{\rm max}} \frac{3}{2} \frac{dN}{d\eta} E\left(\eta\right)d\eta.
\label{eq:inelast}
\end{equation}
with $\eta_{\rm max}=\ln{\left(\sqrt{s}/\mu_{T}\right)}$. 

\begin{figure}
\begin{center}
\includegraphics[width=0.98 \textwidth]{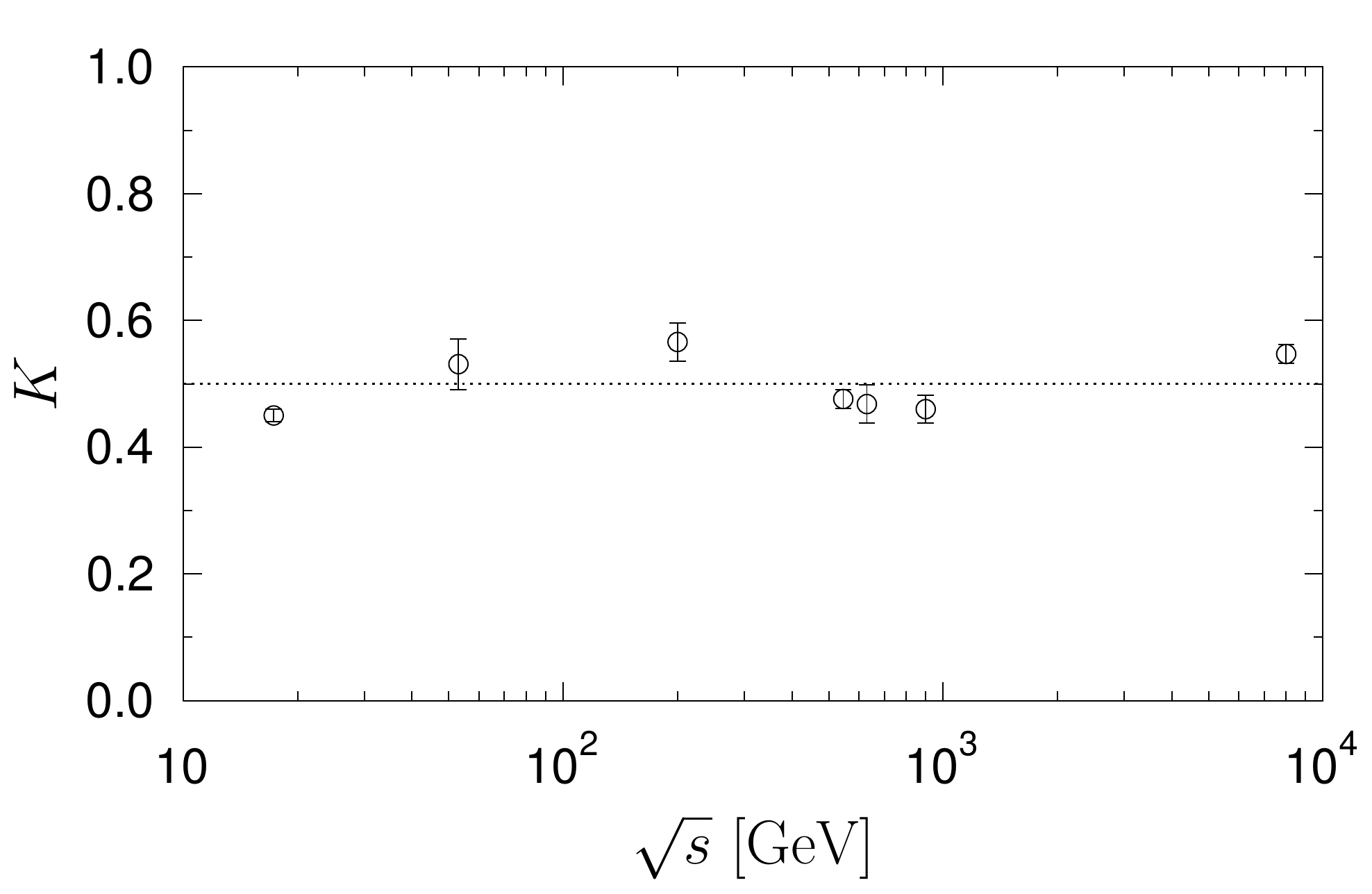} 
\end{center}
\vspace{-5mm}
\caption{Energy dependence of the inelasticity coefficient calculated using fits presented in Fig.~\ref{fig:eta_fits}.} 
\label{fig:K}
\end{figure}

In Fig.~\ref{fig:K} we show the total inelasticity $K=K\left(\sqrt{s}\right)$ obtained by integrating spectra given by Eq.~\ref{eq:dN_eta} with parameters from the fit to the corresponding pseudorapidity distributions as a function of the reaction energy, $\sqrt{s}$.

The overall tendency is such that inelasticity is essentially constant with energy and equal to $K\cong 0.5$, which agrees with first estimates made in~\cite{Cocconi:1958zz} and with first estimates based on the analysis of the leading particle effect provided in~\cite{Basile:1980ap,Basile:1980by}. Over the energy range 17.3~GeV - 8000~GeV, the evaluated inelasticity coefficient varies with dispersion $\sigma=0.05$, and we have $\langle K\rangle=0.50\pm 0.02$. The highest energy data (coming from CMS and TOTEM experiments) do not contradict the results from lower energy data~\cite{Navarra:2003am}.

\section{Concluding remarks}
\label{sec:cr}

Our investigation was aimed at the phenomenological, maximally model independent description, which would eventually result in estimates of inelasticities and their energy dependence. Data for rapidity distributions can be fitted with one free parameter - the longitudinal temperature. The non-extensivity parameter $q$ comes from multiplicity distributions (responsible for dynamical fluctuations existing in hadronizing systems and showing up in the characteristic negative binomial form of the measured {\it multiplicity} distributions) and transverse mass was evaluated from the observed energy dependence of transverse momenta. This means therefore, that in collider data for proton+antiproton collisions and for fixed target proton+proton data analyzed in the same way, there is no additional information to the one used here. From our analysis we can conclude that inelasticity is constant when the collision energy changes by {three} orders of magnitude.


\vspace*{0.3cm}
\centerline{\bf Acknowledgements}
\vspace*{0.3cm}
This research  was supported by the National Science Centre (NCN) grant 2016/23/B/ST2/00692.


\end{document}